\documentclass{article}
\usepackage[utf8]{inputenc} 
\usepackage{graphicx} 
\usepackage{slashed,epsfig,amsmath,amssymb,enumitem} \usepackage[papersize={8.5in,11in}]{geometry}
   \usepackage{bm}
\usepackage{graphicx}
\newcommand{\be}{\begin{equation}}
\newcommand{\ee}{\end{equation}}
\title{Stueckelberg Gauge Invariant Formulation of MOG}
\author{J. W. Moffat\\
Perimeter Institute for Theoretical Physics, Waterloo, Ontario N2L 2Y5, Canada\\
and\\
Department of Physics and Astronomy, University of Waterloo, Waterloo,\\
Ontario N2L 3G1, Canada}
\begin{document}
\maketitle

\begin{abstract} 
We develop a Stueckelberg gauge-invariant formulation of modified gravity (MOG). The massive vector field is made gauge-invariant by introducing a compensating scalar field, without requiring a Higgs field, spontaneous symmetry breaking, or a vacuum expectation value to fix the effective Newtonian gravitational coupling. This separates the gauge-invariant origin of the vector mass from the cosmological evolution of the gravitational coupling. The formulation preserves the finite-range vector interaction of MOG, while allowing the effective gravitational coupling to be treated as an independent scalar or scale-dependent quantity. This distinction is important for cosmological tests, since early-universe constraints and late-time large-scale gravitational phenomena need not be tied to a symmetry-breaking vacuum. The Stueckelberg formulation provides a gauge-invariant framework for comparing MOG with nucleosynthesis, cosmic microwave background, large-scale structure, lensing, and distance data. 
\end{abstract}

\section{Introduction} 
\label{sec:introduction} 

Modified Gravity (MOG), also known as scalar--tensor--vector gravity (STVG), was developed as a relativistic alternative to the dark matter paradigm in which the gravitational interaction is strengthened on galactic and cosmological scales by additional gravitational degrees of freedom~\cite{Moffat2006, BrownsteinMoffat2006,MoffatToth2009, MoffatRahvar2013, MoffatRahvar2014, Moffat2014, Moffat2015,Moffat2016, DavariRahvar, MoffatToth2013, MoffatToth2015,
GreenMoffat2019,BrownsteinMoffat2007,IsraelMoffat,
GreenMoffatToth2018,Moffat2026a,Moffat2026b,Moffat2026c,
Moffat2026d}. The theory contains the metric $g_{\mu\nu}$, a massive vector field $\phi_\mu$, and scalar fields that determine the effective gravitational coupling $G(x)$ and the vector mass scale $\mu(x)$. The vector field mediates a repulsive gravitational charge interaction, while the scalar sector allows the strength and range of the modified force to vary with scale and environment. This structure has been used to describe galaxy rotation curves, cluster dynamics, gravitational lensing, and cosmological observables without introducing nonbaryonic cold dark matter. 

A central theoretical issue in MOG is the status of the massive vector field. A Proca mass term for $\phi_\mu$ gives the vector field a finite range, but it breaks the Abelian gauge invariance. There are two ways to restore gauge invariance. The first is a spontaneous symmetry-breaking formulation~\cite{Sohrab1,Sohrab2}, in which the vector mass and other effective parameters arise from a scalar field with a nonzero vacuum expectation value. The second is the Stueckelberg formulation~\cite{Stueckelberg1938, RueggRuizAltaba2004}, in which a compensating scalar field is introduced so that the massive vector sector remains gauge invariant without requiring a Higgs-type order parameter. These two formulations can be equivalent in a restricted vector sector, but they need not be equivalent when embedded in cosmology. Recent work has investigated a gauge-invariant MOG model based on spontaneous symmetry breaking and has compared its cosmological consequences with observational data ~\cite{Sohrab2}. In that construction, the scalar field associated with the varying gravitational coupling acquires a vacuum expectation value. The effective Newton constant is then tied to the vacuum structure. This identification has important cosmological consequences. If the vacuum expectation value vanishes in a high-temperature phase, gravity is modified or suppressed in the early universe. If the vacuum expectation value evolves or changes across a phase transition, the expansion history, big-bang nucleosynthesis, cosmic microwave background anisotropies, and late-time distance measurements are all affected. The comparison with data is then sensitive not only to the presence of the MOG vector field, but also to the assumed scalar potential and the thermal history of the symmetry-breaking sector. 

The purpose of this paper is to formulate an alternative gauge-invariant version of MOG based on the Stueckelberg mechanism. In this formulation, the massive vector field is made gauge invariant by introducing a compensator field $\sigma$, while the scalar field $G(x)$ is not required to be a vacuum order parameter. 
The Stueckelberg formulation preserves gauge invariance without invoking spontaneous symmetry breaking. The scalar fields $G(x)$ and $\mu(x)$ remain part of the MOG gravitational sector, but their cosmological evolution is determined by the scalar action and by the field equations, not by the requirement that $G$ be generated by a vacuum expectation value. This distinction is physically important. A Stueckelberg formulation separates the gauge-invariant completion of the vector mass from the cosmological origin of the gravitational coupling. The early universe can satisfy the nucleosynthesis requirement:
\begin{equation} 
G_{\rm eff}(k_{\rm BBN},a_{\rm BBN})\simeq G_N , 
\end{equation} 
while the large-scale late-time universe may still exhibit an enhanced effective gravitational coupling:
\begin{equation} G_{\rm eff}(k,a_0)>G_N , 
\end{equation} 
on scales relevant to galaxies, clusters, structure growth, and large-scale velocity flows. This possibility is not automatically available in a formulation in which the same vacuum-order parameter fixes the gravitational coupling in both the early and late universes. The Stueckelberg construction also clarifies the degree-of-freedom content of the theory. In unitary gauge, $\sigma=0$, the vector sector reduces to the usual massive Proca form. In a covariant gauge, the compensator field makes gauge invariance explicit and is useful for perturbation theory, quantization, and cosmological linear-response calculations. Physical differences from a Proca description arise only when the scalar sector or the cosmological evolution of $G_{\rm eff}$ is changed. Physical differences from a spontaneous symmetry breaking formulation arise because no scalar vacuum expectation value is required to generate either the vector mass or the effective Newton constant. 

The paper is organized as follows. In Section 2, we construct the Stueckelberg gauge-invariant action for MOG and derive the field equations for the metric, vector, compensator, and scalar sectors. We specialize the theory to a homogeneous and isotropic cosmological background and identify the modified Friedmann equations. We discuss the linear perturbation equations and the scale-dependent effective gravitational coupling, and describe the implications for nucleosynthesis, cosmic microwave background anisotropies, baryon acoustic oscillations, type-Ia supernovae, and growth data. The conclusions are presented in Section 3.

\section{Stueckelberg Gauge-Invariant Formulation of MOG} 
\label{sec:Stueckelberg_mog} 

The vector sector of scalar--tensor--vector gravity can be formulated in a manifestly gauge-invariant form by introducing a Stueckelberg compensator field. This construction gives the MOG vector field a gauge-invariant mass without requiring a Higgs field, a symmetry-breaking potential, or a vacuum expectation value that fixes the gravitational coupling. It provides an alternative to spontaneous symmetry breaking formulations in which the effective Newton constant is tied to the vacuum value of a scalar order parameter. Let $\phi_\mu$ denote the MOG vector field and define the field strength: 
\begin{equation} 
B_{\mu\nu}=\nabla_\mu \phi_\nu-\nabla_\nu \phi_\mu . 
\end{equation} 
The Stueckelberg scalar $\sigma$ is introduced through the gauge-invariant combination: \begin{equation} 
{\cal A}_\mu = \phi_\mu-\frac{1}{q_\phi}\nabla_\mu\sigma , \label{eq:stueck_combination} 
\end{equation} 
where $q_\phi$ is a dimensional constant and $\sigma$ is a scalar field. The theory is invariant under the local transformation:
\begin{equation} 
\phi_\mu\rightarrow \phi_\mu+\nabla_\mu\chi , \qquad \sigma\rightarrow \sigma+q_\phi \chi , \qquad G\rightarrow G , \qquad \mu\rightarrow \mu , \label{eq:stueck_gauge_transformation} 
\end{equation} 
for an arbitrary scalar gauge function $\chi(x)$. The vector mass parameter $\mu(x)$ may remain a scalar degree of freedom of MOG, but it is not required to arise from a symmetry-breaking vacuum. The massive vector sector can then be written in a locally gauge-invariant form: 
\begin{equation} 
{\cal L}_\phi = -\frac{1}{4}B_{\mu\nu}B^{\mu\nu} -\frac{1}{2}\mu^2(x){\cal A}_\mu{\cal A}^\mu , 
\end{equation} 
without introducing a Higgs field or a spontaneous symmetry breaking vacuum.

A covariant Stueckelberg-MOG action may be written in the form: 
\begin{align} S &= \int d^4x\,\sqrt{-g}\, \bigg[ \frac{1}{16\pi G(x)}\left(R-2\Lambda\right) -\frac{1}{4}B_{\mu\nu}B^{\mu\nu} -\frac{1}{2}\mu^2(x){\cal A}_\mu{\cal A}^\mu \nonumber \\ &\hspace{2.0cm} -\frac{1}{2}K_G(G,\mu)\,\nabla_\alpha G\nabla^\alpha G -\frac{1}{2}K_\mu(G,\mu)\,\nabla_\alpha\mu\nabla^\alpha\mu -U(G,\mu) \bigg] +S_{\rm m}+S_{\rm int}. 
\label{eq:stueck_mog_action} 
\end{align} 
Here, $K_G$ and $K_\mu$ are positive scalar kinetic functions, $U(G,\mu)$ is a general scalar potential, and $S_{\rm m}$ is the matter action. The vector field may couple to a conserved matter current $J^\mu$ through the following:
\begin{equation} 
S_{\rm int} = -\int d^4x\,\sqrt{-g}\,J^\mu\phi_\mu . \label{eq:matter_current_coupling} 
\end{equation} 
The interaction term is gauge invariant up to a boundary term, if we have:
\begin{equation} \nabla_\mu J^\mu=0 . \label{eq:current_conservation} 
\end{equation} 
Equivalently, one may couple the current directly to the invariant field ${\cal A}_\mu$, in which case gauge invariance is manifest. Varying the action with respect to $\phi_\mu$ gives the massive vector field equation:
\begin{equation} \nabla_\nu B^{\nu\mu}-\mu^2{\cal A}^\mu = J^\mu , \label{eq:stueck_vector_equation} 
\end{equation} 
up to the sign convention chosen for $S_{\rm int}$. Variation with respect to $\sigma$ gives the Stueckelberg constraint: 
\begin{equation} \nabla_\mu\left(\mu^2{\cal A}^\mu\right)=0 , \label{eq:stueck_constraint} 
\end{equation} 
in the absence of a direct coupling of the current to ${\cal A}_\mu$. 

Taking the divergence of Eq.~\eqref{eq:stueck_vector_equation} shows that Eq.~\eqref{eq:stueck_constraint} is compatible with current conservation. The Stueckelberg field does not introduce an additional propagating physical polarization beyond the three polarizations of a massive vector field. It restores gauge redundancy and permits the massive vector theory to be treated in a gauge-invariant manner. The metric field equation takes the schematic form:
\begin{equation} \frac{1}{G}G_{\mu\nu} +\frac{\Lambda}{G}g_{\mu\nu} +\left(g_{\mu\nu}\Box-\nabla_\mu\nabla_\nu\right)\frac{1}{G} = 8\pi \left( T^{\rm m}_{\mu\nu} +T^{\phi}_{\mu\nu} +T^{G,\mu}_{\mu\nu} \right), \label{eq:metric_equation_stueck} 
\end{equation} 
where 
\begin{align} T^{\phi}_{\mu\nu} &= B_{\mu\alpha}B_{\nu}{}^{\alpha} -\frac{1}{4}g_{\mu\nu}B_{\alpha\beta}B^{\alpha\beta} +\mu^2 \left( {\cal A}_\mu{\cal A}_\nu -\frac{1}{2}g_{\mu\nu}{\cal A}_\alpha{\cal A}^\alpha \right).
\label{eq:vector_stress_tensor} 
\end{align} 
Here, $T^{G,\mu}_{\mu\nu}$ denotes the stress tensor of the scalar fields $G(x)$ and $\mu(x)$. The scalar equation for $G(x)$ is not fixed by a Stueckelberg gauge principle. It follows from the scalar sector chosen in Eq.~\eqref{eq:stueck_mog_action} and from the nonminimal factor $1/G(x)$ multiplying the Ricci scalar. This is a central difference from a spontaneous symmetry-breaking construction in which $G$ is identified with the vacuum value of a symmetry-breaking field. For a spatially homogeneous and isotropic universe: 
\begin{equation} ds^2=-dt^2+a^2(t)\gamma_{ij}dx^idx^j , 
\end{equation} 
isotropy allows only a temporal background component of the invariant vector: \begin{equation} {\cal A}_\mu dx^\mu={\cal A}_0(t)dt, \qquad B_{\mu\nu}=0. \end{equation} 
The vector equation reduces to an algebraic relation between the temporal vector component and the matter current: 
\begin{equation} 
\mu^2 {\cal A}^0 = -J^0 , \label{eq:cosmo_vector_algebraic} 
\end{equation} 
with the sign depending on the convention used in Eq.~\eqref{eq:matter_current_coupling}. The vector contribution to the homogeneous energy density is given by
\begin{equation} \rho_\phi = \frac{1}{2}\mu^2{\cal A}_0^2 , \label{eq:vector_energy_density} 
\end{equation} 
while the Friedmann equation may be written as:
\begin{equation} H^2 = \frac{8\pi G(t)}{3} \left( \rho_{\rm m} +\rho_{\rm r} +\rho_\phi +\rho_{G,\mu} \right) +\frac{\Lambda}{3} -H\,\frac{\dot G}{G} -\frac{k}{a^2} . \label{eq:stueck_friedmann} 
\end{equation} 
The term proportional to $\dot G/G$ arises from the variation of the non-minimal factor $1/G(t)$ in the gravitational action, and $k=0,1,-1$, for a flat, closed and open universe. Its detailed form depends on the normalization of the scalar kinetic sector. 

The Stueckelberg formulation separates two logically distinct ingredients: the gauge-invariant mass of the vector field and the cosmological evolution of the gravitational coupling. Gauge invariance of the massive vector field does not require:
\begin{equation} 
G_{\rm eff}\propto \langle \Phi^\dagger\Phi\rangle , \label{eq:no_g_vev_requirement} 
\end{equation} 
nor does it require a finite-temperature phase transition in which gravity is absent in the symmetric phase and generated in the broken phase. The function $G(x)$ may instead be treated as an independent scalar field, as a slowly varying cosmological background, or as the local representation of a scale-dependent effective coupling: 
\begin{equation} 
G_{\rm eff}=G_{\rm eff}(k,a) . 
\label{eq:scale_dependent_geff} 
\end{equation} 

In the weak-field cosmological limit of STVG--MOG, the gravitational response of a Fourier mode is governed by a scale- and time-dependent effective Newton constant~\cite{Moffat2026c,Moffat2026d}. Writing the physical wavenumber as $q=\frac{k}{a}$, the modified Poisson equation may be expressed as: 
\begin{equation}
k^2\Phi(k,a) = -4\pi G_{\rm eff}(k,a)a^2\rho(a)\delta(k,a), 
\end{equation}
where $\delta({\bf x},a) \equiv \frac{\rho({\bf x},a)-\bar{\rho}(a)}{\bar{\rho}(a)} = \frac{\delta\rho({\bf x},a)}{\bar{\rho}(a)}$, where \(\bar{\rho}(a)\) is the homogeneous background density. We have the following:
\begin{equation}
G_{\rm eff}(k,a) = G_N\left[1+\alpha_{\rm eff}(k,a)\right], \qquad \alpha_{\rm eff}(k,a) = \alpha\,\frac{\mu^2}{k^2/a^2+\mu^2}. 
\end{equation}
Equivalently: 
\begin{equation}
G_{\rm eff}(k,a) = G_N\left[1+\alpha\,\frac{\mu^2}{q^2+\mu^2}\right]. 
\end{equation}
Here, \(G_N\) is Newton's constant, \(\alpha\) measures the strength of the MOG enhancement, and \(\mu^{-1}\) is the range associated with the vector-field Yukawa sector. For modes with \(q\gg \mu\), the Yukawa screening suppresses the modification and we have:
\begin{equation}
G_{\rm eff}(k,a)\simeq G_N . 
\end{equation}
For modes with \(q\lesssim \mu\), the screening becomes ineffective and the coupling approaches 
\begin{equation}
G_{\rm eff}(k,a)\simeq G_N(1+\alpha). 
\end{equation}
This behavior allows STVG--MOG to reproduce the gravitational response required by isotropic linear observables on the relevant scales, while leaving a distinctive large-scale regime in which anisotropic observables, such as radio-galaxy and quasar number-count dipoles, can probe departures from the \(\Lambda\)CDM limit.

The distinction $G_{\rm eff}(k,a)$ is important for cosmological tests. In a symmetry-breaking formulation, the same vacuum order parameter that generates the gravitational coupling can affect early-universe expansion, nucleosynthesis, late-time acceleration, and the relation between local and cosmological measurements of Newton's constant. In a Stueckelberg formulation, those issues are not tied to the gauge-invariant completion of the vector mass. The early-universe constraint may be imposed as given by
\begin{equation} G_{\rm eff}(k_{\rm BBN},a_{\rm BBN})\simeq G_N,
\label{eq:bbn_limit} 
\end{equation} 
while the late-time large-scale regime allows: 
\begin{equation} G_{\rm eff}(k,a_0)>G_N , \qquad k/a_0 \lesssim \mu , \label{eq:late_time_geff} 
\end{equation} 
without invoking a vacuum transition for $G$. The Stueckelberg construction also clarifies the physical degree-of-freedom count. In unitary gauge, $\sigma=0$, the invariant field ${\cal A}_\mu$ reduces to $\phi_\mu$, and the vector sector becomes the usual massive Proca form. In a covariant gauge, the compensator field keeps gauge invariance explicit and can simplify perturbation theory. Observable differences from a Proca or Higgs-like formulation arise only if the scalar sector, the cosmological evolution of $G$, or the relation between $\mu$ and $G_{\rm eff}$ is changed. A pure Stueckelberg rewriting of the same massive vector theory is physically equivalent to the Proca theory after gauge fixing. 

A Stueckelberg-MOG theory with an independent $G(x)$ sector is not equivalent to a spontaneous-symmetry-breaking theory in which $G$ is fixed by a vacuum expectation value. For comparison with cosmological data, the field equations following from Eq.~\eqref{eq:stueck_mog_action} should be evolved at the background and linear perturbation levels. The relevant observables include the nucleosynthesis expansion rate, the acoustic scale of the cosmic microwave background, the growth function $D_+(a)$, weak-lensing amplitudes, baryon acoustic oscillations, type-Ia supernova distances, and large-scale velocity flows. The Stueckelberg version is especially useful because it permits the massive MOG vector field to remain gauge invariant, while leaving open the phenomenologically important question of whether $G_{\rm eff}$ is constant, time dependent, or scale dependent.

\section{Conclusions} 
\label{sec:conclusions} 

The principal result of the gauge-invariant version of MOG based on the Stueckelberg mechanism is the separation of two issues that are often tied together in symmetry-breaking versions of gauge-invariant MOG: the gauge-invariant origin of the vector mass and the cosmological origin of the effective gravitational coupling. In the Stueckelberg theory the vector field can be massive and gauge invariant without requiring the Newton coupling $G(x)$ to be generated by a scalar vacuum expectation value. The scalar field $G(x)$ may instead be treated as an independent gravitational degree of freedom, as a slowly varying cosmological background, or as a local description of a scale-dependent effective coupling: $G_{\rm eff}=G_{\rm eff}(k,a)$. This distinction is important for cosmology because a theory in which $G_{\rm eff}$ is fixed by a vacuum order parameter can impose strong constraints on the early universe, the thermal history, and the relation between big-bang nucleosynthesis and late-time large-scale structure. The Stueckelberg formulation keeps the physical degree of freedom content of the massive vector sector under control. 

In unitary gauge, $\sigma=0$, the theory reduces to the Proca form used in the original MOG action. In a covariant gauge, the compensator field preserves gauge invariance and is useful for perturbation theory and quantization. The Stueckelberg field does not introduce an additional physical polarization beyond the three polarizations of the massive vector field. Observable differences arise only when the scalar sector, the cosmological evolution of $G(x)$, or the scale-dependence of $G_{\rm eff}$ differs from the corresponding Proca or spontaneous-symmetry-breaking theory. For homogeneous and isotropic cosmology, the background vector field is restricted by spatial isotropy to a temporal component. The vector equation then gives an algebraic relation between the temporal invariant vector component and the matter current. 

The Friedmann equations are modified by the scalar dynamics of $G(x)$ and $\mu(x)$, by the energy density of the vector field, and by the derivative terms generated by the non-minimal coupling $1/G(x)$ multiplying the Ricci scalar. The resulting background cosmology can be compared with big-bang nucleosynthesis, type-Ia supernovae, cosmic chronometers, baryon acoustic oscillations, and the cosmic microwave background. The Stueckelberg version also provides a natural setting for a scale-dependent gravitational coupling. The early-universe constraint may be implemented as:
\begin{equation} G_{\rm eff}(k,a)\simeq G_N , 
\end{equation} 
while the late-time large-scale regime may allow: 
\begin{equation} G_{\rm eff}(k,a_0)>G_N,
\end{equation} 
on scales relevant to galaxies, clusters, structure growth, lensing, and bulk flows. 

This possibility is not forced by the Stueckelberg mechanism itself, but the formulation permits it without associating the change in gravitational strength with a vacuum phase transition. It also avoids the interpretation that gravity is absent or strongly suppressed in a symmetric high-temperature phase. The comparison between the Stueckelberg and spontaneous-symmetry-breaking formulations can be summarized as follows. A pure Stueckelberg rewriting of a massive Abelian vector theory is physically equivalent to the Proca theory after gauge fixing. A spontaneous-symmetry-breaking formulation introduces a physical scalar order parameter and a scalar potential. If the gravitational coupling is identified with the vacuum expectation value of that scalar, the cosmological evolution of $G_{\rm eff}$ becomes tied to the vacuum structure of the theory. A Stueckelberg-MOG formulation with an independent $G(x)$ sector does not make this identification and can lead to different cosmological predictions. A complete confrontation with data requires deriving and numerically integrating the linear scalar, vector, and tensor perturbation equations. The quantities to be computed include the effective Poisson equation, the gravitational slip, the growth function $D_+(a)$, the lensing potential, the CMB temperature and polarization spectra, baryon-acoustic-oscillation distances, weak-lensing amplitudes, and peculiar-velocity observables. These calculations will determine whether a Stueckelberg gauge-invariant MOG theory can satisfy early-universe constraints, while retaining the successful large-scale phenomenology associated with an enhanced gravitational coupling. 

The Stueckelberg formulation offers a gauge-invariant completion of the MOG vector sector. It preserves the massive finite-range vector interaction, maintains explicit gauge invariance, and does not require the Newtonian gravitational constant to arise from a symmetry-breaking vacuum. For this reason, it provides a useful framework for testing whether the cosmological successes and tensions of gauge-invariant MOG are consequences of the vector field itself, or consequences of the additional assumption that the gravitational coupling is fixed by spontaneous symmetry breaking.

\section*{Acknowledgments}

We thank Viktor Toth and Martin Green for helpful discussions. Research at the Perimeter Institute for Theoretical Physics is supported by the Government of Canada through Industry Canada and by the Province of Ontario through the Ministry of Research and Innovation (MRI).

\end{document}